\documentclass[preprint,twocolumn]{rmaa}
\usepackage{natbib}
\usepackage{paralist}
\usepackage{graphicx} 
\usepackage[usenames]{color}
\usepackage{psfrag,color}  
\usepackage{ulem}
\usepackage[latin1]{inputenc}

\title{The cometary cavity created by an aligned
streaming environment/collimated outflow interaction}

\author{D. L\'opez-C\'amara\altaffilmark{1}, A. Esquivel\altaffilmark{1},
J. Cant\'o\altaffilmark{2}, A. C. Raga\altaffilmark{1}, P. F Vel\'azquez\altaffilmark{1}, \& A. Rodr{\'{\i}}guez-Gonz{\'a}lez\altaffilmark{1}}

\altaffiltext{1}{Instituto de Ciencias Nucleares, Universidad Nacional Aut\'onoma de M\'exico, Ap. 70-543, 04510 D.F., M\'exico}

\altaffiltext{2}{Instituto de Astronom\'\i a, Universidad Nacional Aut\'onoma de M\'exico, Ap. 70-468, 04510 D.F., M\'exico}

\shortauthor{L\'opez-C\'amara et al.}
\shorttitle{Jet streaming environment cavities}

\fulladdresses{
\item D. L\'opez-C\'amara, A. Esquivel, A. C. Raga, P. F. Vel\'azquez,
A. Rodr\'\i guez-Gonz\'alez: Instituto de Ciencias Nucleares,
Universidad Nacional Aut\'onoma de M\'exico, Ap. 70-543, 04510  D.F.,
M\'exico 
(\email{d.lopez, esquivel, raga, pablo, ary@nucleares.unam.mx})
\item J. Cant\'o: Instituto de Astronom\'\i a, Universidad Nacional
Aut\'onoma de M\'exico, Ap. 70-264, 04510  D.F., M\'exico                       
}                                  

\listofauthors{D. L\'opez-C\'amara, A. Esquivel, J., Cant\'o,                   
A. C. Raga, P. F. Vel\'azquez, A. Rodr\'\i guez-Gonz\'alez}

\indexauthor{L\'opez-C\'amara, D.}
\indexauthor{Esquivel, A.}
\indexauthor{Cant\'o, J.}
\indexauthor{Raga, A. C.}
\indexauthor{Vel\'azquez, P. F.}
\indexauthor{Rodr\'\i guez-Gonz\'alez, A.}

\abstract{
We present a ``thin shell'' model of the interaction of a biconical
outflow and a streaming environment (aligned with the direction of the
flow), as well as numerical (axisymmetric) simulations of such an
interaction. A similar situation, although in a more complex setup,
takes place at the head of the cometary structure of Mira. Thus, for
most of the numerical simulations we explore parameters consistent
with the observed bipolar outflow from Mira B. For these
parameters, the interaction is non-radiative, so that a rather broad
jet/streaming environment interaction region is formed. In spite
of this, a reasonable agreement between the thin-shell analytic model and
the numerical simulations is obtained.
}

\resumen{
Presentamos un modelo anal\'{\i}tico de ``capa delgada'' de la
interacci\'on de un chorro bic\'onico y un medio ambiente en movimiento
(alineado con la direcci\'on del chorro), as\'{\i} como simulaciones
num\'ericas (axisim\'etricas) de dicha interacci\'on. 
Una situaci\'on similar, aunque en un escenario m\'as complejo, ocurre
en la cabeza de la estructura cometaria de Mira. Por esta raz\'on, en
la mayor\'{\i}a de las simulaciones num\'ericas exploramos par\'ametros
que son consistentes con las observaciones del flujo bipolar de Mira B.
Para estos par\'ametros  la interacci\'on es
no-radiativa, lo que resulta en una zona de interacci\'on jet/medio
ambiente bastante ancha. A pesar de esto,
encontramos que el modelo anal\'\i tico de capa delgada
describe de manera satisfactoria la
morfolog\'{\i}a b\'asica del flujo.
}
\addkeyword{hydrodynamics}
\addkeyword{ISM: dynamics}
\addkeyword{ISM: Jets and outflows}
\addkeyword{Stars: AGB and post-AGB}
\addkeyword{Stars: Mass loss}

\begin{document}

\maketitle

\section{Introduction}
\label{sec:int}

Prompted by the discovery of a striking $\sim 2\arcdeg$ cometary tail
attached to Mira by \citet{m07}, a series of theoretical and
observational studies of this object have been made.
Theoretical works include \citet{w07, r08, rc08, esquivel10}, all of
which are devoted to study the interaction between the post-AGB wind (from
Mira A) and a streaming interstellar medium (ISM). This streaming medium
corresponds to the motion of Mira through the galactic plane.
Observations of the cometary tail have been done in several wavebands,
including  21cm \citep{m08} and the IR \citep{u08}.

Optical (H$\alpha$) observations by \citet{m09} revealed a fast bipolar
outflow, which probably arises from the less luminous component, Mira B.
This outflow is embedded in the head of the cometary structure.
It has radial velocities of $\pm 150$~km~s$^{-1}$
(with respect to Mira), is relatively broad, and its projection onto
the plane of the sky is approximately aligned with the direction of
Mira's proper motion.

The mass loss rate from Mira A has estimated values from
$10^{-7}$ to $10^{-5}$
M$_{\odot}$yr$^{-1}$ \citep{km85,bk88}. Part of
this mass loss is accreted onto a disk around Mira B (and eventually
onto Mira B itself). Ireland et al. (2007) estimated a $\sim 5\times
10^{-9}$ M$_{\odot}$yr$^{-1}$ accretion rate. Part of this accretion
rate would then be redirected into the bipolar outflow from Mira B.

In the Mira A/B system, we therefore appear to have a bipolar
outflow from Mira B which first interacts with the wind from
Mira A and at larger distances emerges from the wind into
the streaming ISM region. However, in the present paper we
consider the more simple problem of an aligned (bi-conical)
bipolar outflow/streaming environment interaction (i. e.,
neglecting the presence of the wind from Mira A). This is
a first step to modeling the high velocity outflow observed
by \citet{m09}.

We present an analytic, stationary, thin shell model
of the interaction of a bi-conical outflow and an aligned, streaming
environment in Section~\ref{sec:anal}.
Time-dependent, axisymmetric numerical simulations of this problem
are described in Section~\ref{sec:simul}, and their results presented
in Section~\ref{sec:results}. With the simulations we show the
characteristics of the flow in greater detail, and evaluate the level
of agreement with the analytic, thin shell solution. Finally,
the implications of this model for the interpretation of
Mira's outflow are discussed in Section~\ref{sec:disc}.

\section{The analytic model} 
\label{sec:anal}

\begin{figure}[!h]
\includegraphics[width=\columnwidth]{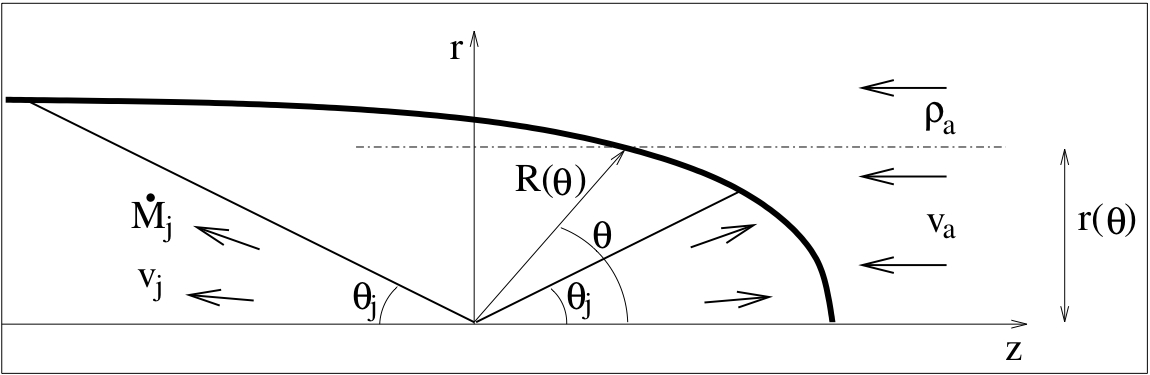}
\caption{Schematic diagram showing the interaction between a
  bi-conical outflow (of half-opening angle $\theta_j$, velocity $v_j$
  and total mass loss rate $2{\dot M}_j$) with a streaming environment
  of density $\rho_a$ and velocity $v_a$ (aligned with the outflow
  axis).}
\label{fig1}
\end{figure}

Let us consider a source that ejects a bipolar, conical
outflow (with a half-opening angle $\theta_j$, velocity $v_j$
and mass loss rate ${\dot M}_j$ for each outflow lobe), immersed
in a uniform, streaming environment of density $\rho_a$,
which travels parallel to the outflow axis at a velocity $v_a$.
The situation is shown in the schematic diagram of Figure \ref{fig1}.
We assume that as the material from the outflow encounters the
streaming environment a thin shell of well mixed material is formed
(the thick solid line shown in Fig \ref{fig1}). This layer has a bow
shock shape $R(\theta)$, where $R$ is the spherical radius and
$\theta$ the angle measured from the $z$-axis, see Figure 1).

At a position $\theta$ along the bow shock, the material flowing
along the thin shell has a mass rate:
\begin{equation}
{\dot M}(\theta)=\int_0^\theta d{\dot M}_j+\pi r^2\rho_a v_a\,,
\label{m}
\end{equation}
where the first term on the right represents the mass fed into
the shell by the conical jet, and the second term is the mass from 
the streaming environment (with $r=R\sin\theta$ being
the cylindrical radius, see Figure \ref{fig1}).

The radial momentum rate of the material in the shell is:
\begin{equation}
{\dot \Pi}_r(\theta)={\dot M}(\theta)\,v_r(\theta)=
\int_0^\theta v_j\sin\theta\,d{\dot M}_j\,,
\label{pir}
\end{equation}
with a contribution only from the jet. The axial momentum (along the
$z$-direction) flowing along the shell is: 
\begin{equation}
{\dot \Pi}_z(\theta)={\dot M}(\theta)\,v_z(\theta)=
\int_0^\theta v_j\cos\theta\,d{\dot M}_j-\pi r^2\rho_a v_a^2\,.
\label{piz}
\end{equation}
In equations (\ref{pir}) and (\ref{piz}), 
$v_r$ and $v_z$ are the $r$- and $z$-components of the velocity
of the (well mixed) material flowing along the thin shell.
One can also write an expression for the angular momentum rate of the
material flowing along the shell:
$$
{\dot J}(\theta)={\dot M}(\theta)\,v_\theta(\theta)R(\theta)=
\int_0^r r'v_a(2\pi 4'\rho_av_a)dr'
$$
\begin{equation}
=\frac{2}{3}\pi \rho_a
v_a^2R^3\sin^3\theta\,,
\label{j}
\end{equation}
where  $v_\theta$ is the velocity in the $\theta$-direction (of
the material flowing along the thin shell), which is given by:
\begin{equation}
v_\theta=v_r\cos\theta-v_z\sin\theta\,.
\label{vth}
\end{equation}

Now, we consider a  conical, bipolar outflow with a mass loss rate given by:
\begin{equation}
  d{\dot M}_j=\left\{ \begin{array}{cl}
    \frac{\dot M_j}{{1-\cos\theta_j}}\sin\theta\, d\theta\,
    &;\ \theta\leq \theta_j\,\,{\rm or}\,\,\theta\geq
    \pi-\theta_j\,,\\
    0
    &;\ \theta_j<\theta<\pi-\theta_j\,.
  \end{array} \right.
\label{dm}
\end{equation}
With this form for $d{\dot M}_j$, it is possible to carry out the
integrals in equations (\ref{m}-\ref{piz}), allowing to obtain the
shape of the bow shock: 
\begin{equation}
R(\theta)=R_0\,{\rm cosec\,}\theta\times\sqrt{3\left(1-\theta\,{\rm cotan\,}
\theta\right)}\,; \,\,\,\theta\leq\theta_j\,,
\label{r1}
\end{equation}
$$
R(\theta)=R_0\,{\rm cosec\,}\theta
$$
$$
\times \sqrt{3\left[\cos(\theta-\theta_j)
\sin\theta_j\,{\rm cosec\,}\theta -\theta_j\,{\rm cotan\,}
\theta\right]}\,;
$$
\begin{equation}
\,\,\,\theta_j<\theta<\pi-\theta_j\,,
\label{r2}
\end{equation}
$$
R(\theta)=R_0\,{\rm cosec\,}\theta
$$
$$
\times \sqrt{3\left[1+
\left(\pi-\theta-2\theta_j+\sin 2\theta_j\right)\,{\rm cotan\,}
\theta\right]}\,;
$$
\begin{equation}
\,\,\,\pi-\theta_j\leq \theta\leq\pi\,,
\label{r3}
\end{equation}
where
\begin{equation}
R_0\equiv \sqrt{\frac{{\dot M}_jv_j}{{2\pi(1-\cos\theta_j)\rho_a
v_a^2}}}\,,
\label{r0}
\end{equation}
is the on-axis standoff distance between the bow shock and the
jet source.

\begin{figure}[!h]
\includegraphics[width=1.0\columnwidth]{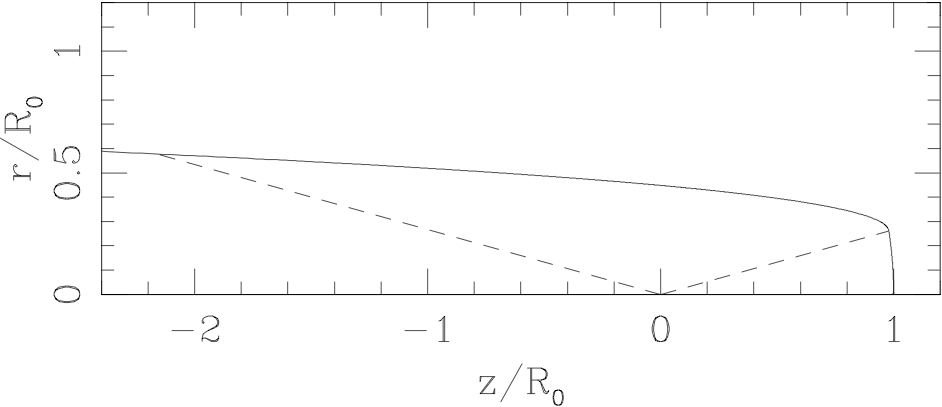}
\caption{Analytic solution for the shape of the bow shock
produced by the interaction of a bi-conical outflow with an
aligned, streaming environment. The outflow cones have
a half-opening angle $\theta_j=15^\circ$, and their outer
boundaries are shown by the dashed lines.}
\label{fig2}
\end{figure}

As an example, in Figure \ref{fig2} we show the bow shock shape
obtained from equations (\ref{r1}-\ref{r3}) for a biconical
outflow with a half-opening angle $\theta_j=15^\circ$. The
solution is quite flat-topped in the $\theta<\theta_j$ ``head''
region. For $\theta>\theta_j$ the solution first rapidly curves,
and then straightens out into a slowly expanding bow shock wing.

\section{The numerical simulations}
\label{sec:simul}

We have computed seven different axisymmetric gasdynamic simulations of
aligned bi-conical jet/streaming environment interactions with the 
{\sc walicxe} code, which is described in detail by \citet{esquivel10}.
The code integrates the gasdynamic equations with an approximate
Riemann solver (in this case we used a hybrid HLL-HLLC algorithm, see
\citealt{hll,hllc,esquivel10}). A hydrogen ionization
rate equation is solved along with the gasdynamic equations in order
to include the radiative losses through a parametrized cooling function,
that depends on the density, temperature and hydrogen ionization fraction
\citep{rr04}.

The {\sc walicxe} code has a block based adaptive grid, with blocks of
a constant number of cells (in our case, of $24\times 24$  cells),
which are refined by successive factors of 2. 
Most of our simulations were run at a medium resolution
(labeled ``mr'' in Table 1). These consist of two root blocks aligned
in the axial direction, which are allowed to have 8 levels of refinement.
The resolution at the finest level would correspond to $6144\times
3072$ cells (axial and radial, respectively) in a uniform grid.
In addition, we ran low and high resolution versions of model 1
(labeled ``lr'' and ``hr'' in Table 1).
The low resolution run has 7 levels of refinement ($3072\times
1536$ cells in a uniform grid), while the high resolution run has 9
levels of refinement ($12288\times 6144$ cells in a uniform
grid).

Models 1-6 have a computational box of the same size, $[2\times
4]10^{18}~\mathrm{cm}$ (radial and axial). Therefore the ``mr'' runs
have a maximum resolution of $6.51 \times 10^{14}$~cm, and the
``lr'' and ``hr'' runs have resolutions of  $1.30 \times 10^{15}$~cm
and  $3.25 \times 10^{14}$~cm, respectively.
In model 7 we were more interested in resolving the
structure at the head of the shock than in the long cometary tail.
To achieve a higher resolution  ($6.51 \times 10^{13}$~cm) with the
same number of computational cells, in these models we have reduced
the size of the computational box by a factor of $10$.

The simulations were performed in a cylindrical computational grid
with an axial extent $[z_{min},z_{max}]$ (with the jet source placed
at $z=0$) and extending radially from the symmetry axis ($r=0$, where
a reflecting boundary condition is applied) out to $r_{max}$. An
inflow condition is applied in the positive $z$ grid 
boundary, where a streaming environment is injected with a constant
velocity in the $-z$ direction. Outflow boundaries were applied at the
negative $z$ grid boundary and at $r_{max}$.
In models 1-3
$z_{min}=-2.8\times 10^{18}~\mathrm{cm}$ and 
$z_{max}= 1.2\times 10^{18}~\mathrm{cm}$, in models 4-6 
$z_{min}=-2.0\times 10^{18}~\mathrm{cm}$ and 
$z_{max}= 2.0\times 10^{18}~\mathrm{cm}$, in model 7
$z_{min}=-2.8\times 10^{17}~\mathrm{cm}$ and
$z_{max}=1.2\times 10^{17}~\mathrm{cm}$.
 
The parameters of the simulations are summarized in Table 1.
The jets are imposed in two cones with a half-opening angle $\Theta$
and an initial outer radius equal to $R_j=10^{16}$~cm for models 1-6,
and $R_j=10^{15}$~cm for model 7. The outflows have a velocity
$v_j$ (see Table 1) and a temperature $T_j=100$~K.
The density inside the source follows a $\propto R^{-2}$ profile set
by the mass loss rate ${\dot M}_j$.
The environment has an initial temperature
$T_a=10^4$~K, a density $n_a$ and a velocity $v_a$. The values for the
parameters of models 1-6 are roughly consistent with the jet of Mira
B while model 7 (with a denser streaming environment)
was chosen to produce a more radiative interaction region.
Both the jet and the environment are initially
neutral, except for a seed electron density assumed to come
from the presence of singly ionized carbon.

In Table 1 we have also included the corresponding value of the
bow-shock standoff distance $R_0$ (see equation \ref{r0}),
and the cooling distances
associated with the double shock structure at $R_0$. The cooling
distance ${d_c}^{f}$ corresponds to the forward shock (with the streaming
ambient medium), while ${d_c}^{r}$ corresponds to the cooling behind
the reverse shock.
In order to calculate these cooling distances, we have used the fit:
$${d_c}^{i}=\left(\frac{\rm 100\,cm^{-3}}{n_0^{i}}\right)
\Biggl\{\left[{3\times 10^{11}\rm cm}\right]
\left(\frac{u_0^{i}}{\rm 100\,km\,s^{-1}}\right)^{-6.4}
$$
\begin{equation}
+\left[{8\times 10^{13}\rm cm}\right]
\left(\frac{u_0^{i}}{\rm 100\,km\,s^{-1}}\right)^{5.5}\Biggr\}\,,
\label{dc}
\end{equation}
to the cooling distances to $10^4$~K given in the tabulation
of self-consistent preionization, plane-parallel shock models of
\citet{har87}. In equation (\ref{dc}), $n_0$ is the pre-shock atom+ion
number density, and $u_0$ is the shock velocity.

\begin{table*}[!t]\centering
\caption{Model characteristics.}
\label{tab1}
\footnotesize{
\begin{tabular}{cccccccccc}
\hline
\hline
Model &  $\Theta$ & $v_j$ & $\dot{M_j}$ & $v_a$ & $n_a$ & $R_0$ & ${d_c}^{r} / (R_0$tan$\Theta)$ & ${d_c}^{f} / (R_0$tan$\Theta)$ & resolution\\
           &   [$^{\circ}$]  & [km s$^{-1}$] & [$M_{\odot}$ s$^{-1}$] & [km s$^{-1}$] & [cm$^{-3}$] & [$10^{17}$~cm] &  & \\
\hline
{\bf 1} & 10 & 200  & 1.00 $\times$10$^{-10}$ & 125 & 0.044 & \phantom{0}2.97 & 402.4 & 11.9 & lr,mr,hr\\
{\bf 2} & 20 &200   & 3.97$\times$10$^{-10}$ & 125 & 0.044 &  \phantom{0}2.97 & 194.9 & \phantom{0}5.7 & mr \\
{\bf 3} & 30 &200   & 8.82$\times$10$^{-10}$ & 125 & 0.044 &  \phantom{0}2.97 & 122.9 & \phantom{0}3.6 & mr \\
{\bf 4} & 10 &200   & 1.00 $\times$10$^{-9}\phantom{0}$ & 125 & 0.044 &  \phantom{0}9.39 & 127.2 & \phantom{0}3.7 & mr \\
{\bf 5} & 20 & 200  & 3.97$\times$10$^{-9\phantom{0}}$ & 125 & 0.044 &  \phantom{0}9.39  & \phantom{0}61.6 & \phantom{0}1.8 & mr \\
{\bf 6} & 30 & 200  & 8.82$\times$10$^{-9\phantom{0}}$ & 125 & 0.044 &  \phantom{0}9.39 & \phantom{0}38.9 & \phantom{0}1.1  & mr \\
{\bf 7} & 10 & \phantom{0}40 & 1.00$\times$10$^{-10\phantom{0}}$ & 100 & 1.0\phantom{0}\phantom{0} &  \phantom{0}3.48 & \phantom{0}0.28 & \phantom{0}1.3 & mr \\
\hline
\hline
\end{tabular}
}
\end{table*}

\section{Results}
\label{sec:results}

\begin{figure}[!h]
\includegraphics[width=1.0\columnwidth]{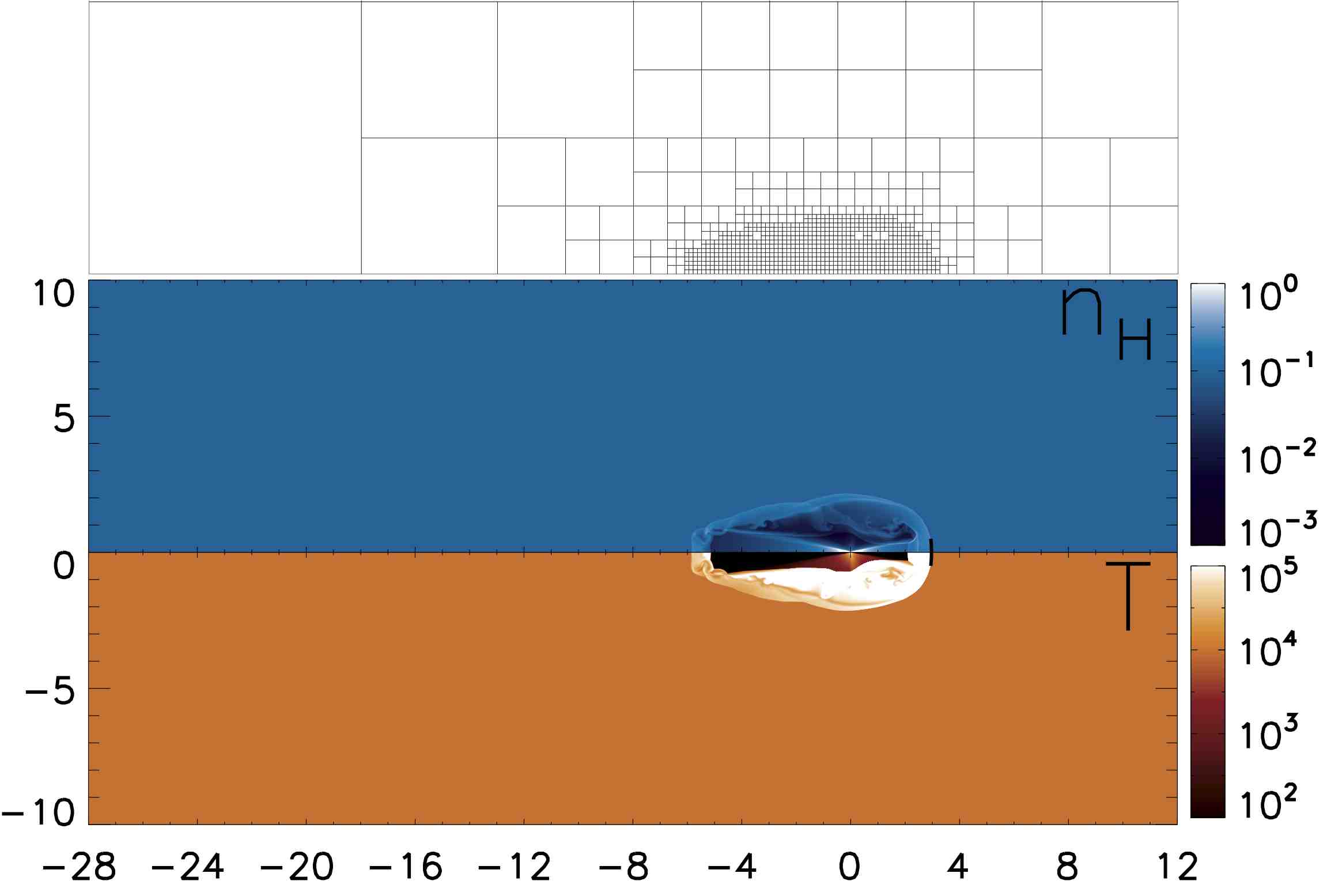}
\caption{Density (color scale given in cm$^{-3}$, middle panel) and temperature
(color scale given in K, bottom panel) stratifications. The corresponding mesh configurations (on top of each density/temperature plot) are also shown, with each square representing a $24\times 24$ grid point block. The results are for model 1 at $t=1\times10^{3}$~yr. The axes are labeled in units of $10^{17}$~cm (the figure only shows the inner half of the radial extent, the rest is unperturbed.).}
\label{fig3a}
\end{figure}

\begin{figure}[!h]
\includegraphics[width=1.0\columnwidth]{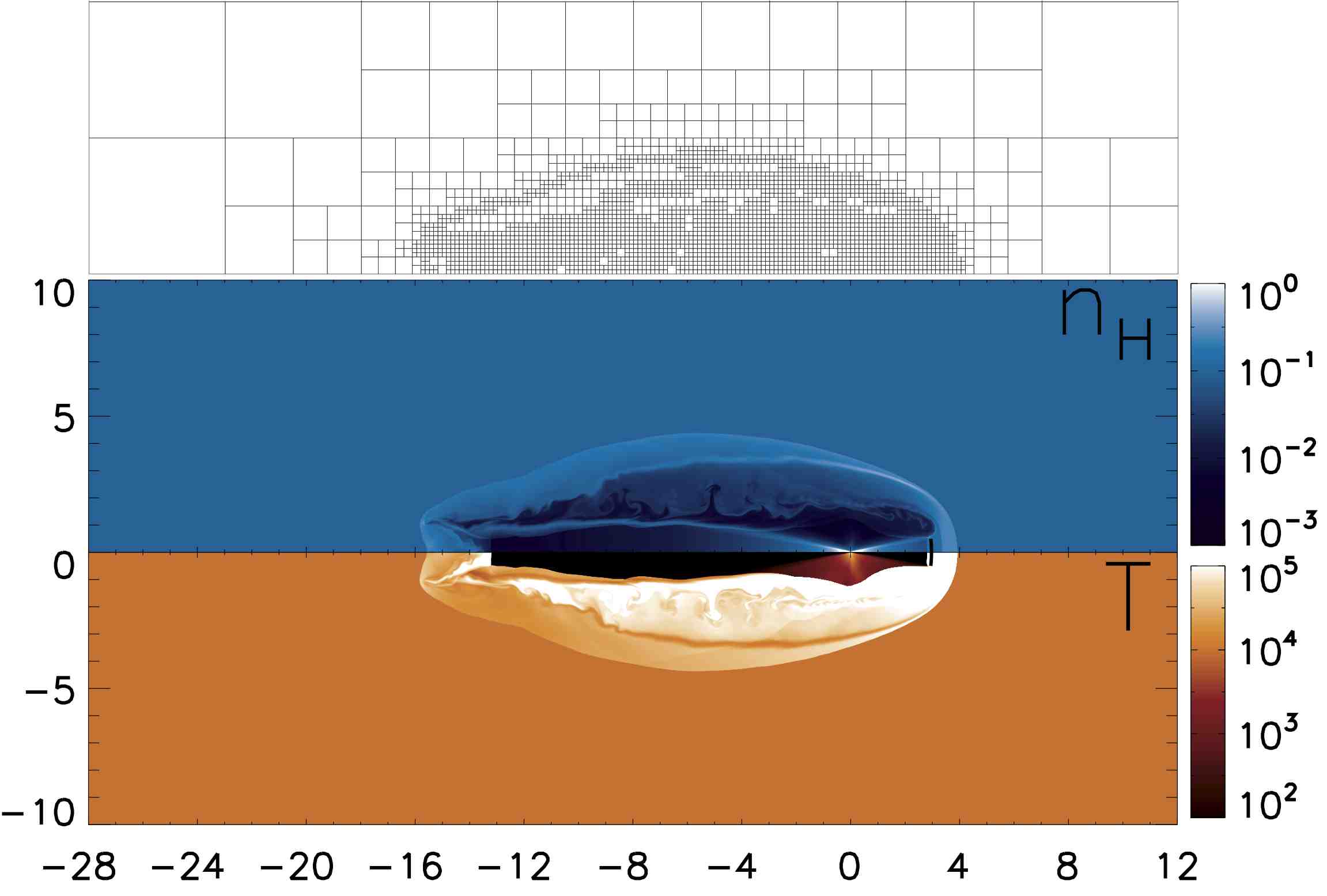}
\caption{Same as Figure~\ref{fig3a} at $t=3\times10^{3}$~yr.}
\label{fig3b}
\end{figure}

\begin{figure}[!h]
\includegraphics[width=1.0\columnwidth]{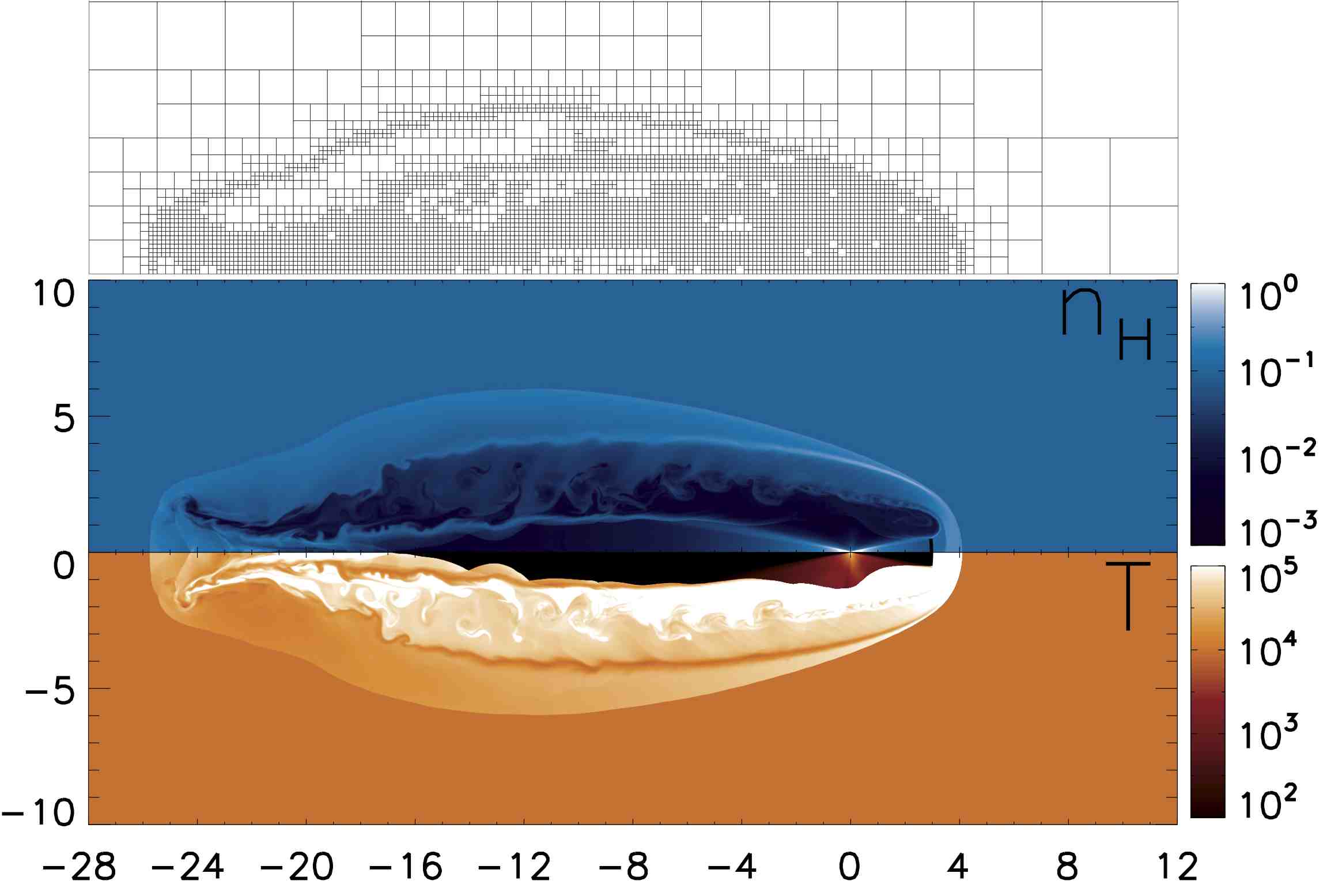}
\caption{Same as Figure~\ref{fig3a} at $t=5\times10^{3}$~yr.}
\label{fig3c}
\end{figure}

In Figures~\ref{fig3a} - \ref{fig3c}, we show the time-evolution ($t=1\times10^{3}$,
3$\times10^{3}$, and 5$\times10^{3}$~yr, respectively) obtained from
model 1 (see Table~1). For each time, we show the corresponding mesh
configuration, the density $n_H$ (in cm$^{-3}$) and the
temperature $T$ (in K) (top, middle, and bottom panels, respectively).
Thanks to the adaptive grid of the code we used
less than 10\% (for $t\times10^{3}$~yr) and 20\% (for
$t>10^{3}$~yr) of the finest grid in the domain, thus reducing notably
the required computer time.

From the time-sequence shown in Figures~\ref{fig3a} - \ref{fig3c}, we see
that the flow develops a cometary-shaped structure, due to the
presence of the side-streaming environmental wind (which
flows parallel to the $z$-axis, from right to left).
The head of the cometary structure approaches the
steady state star/bow shock stagnation distance
$R_0$ (see \S 2) at a time $t \approx 10^{3}$~yr.

\begin{figure}[!h]
\includegraphics[width=1.0\columnwidth]{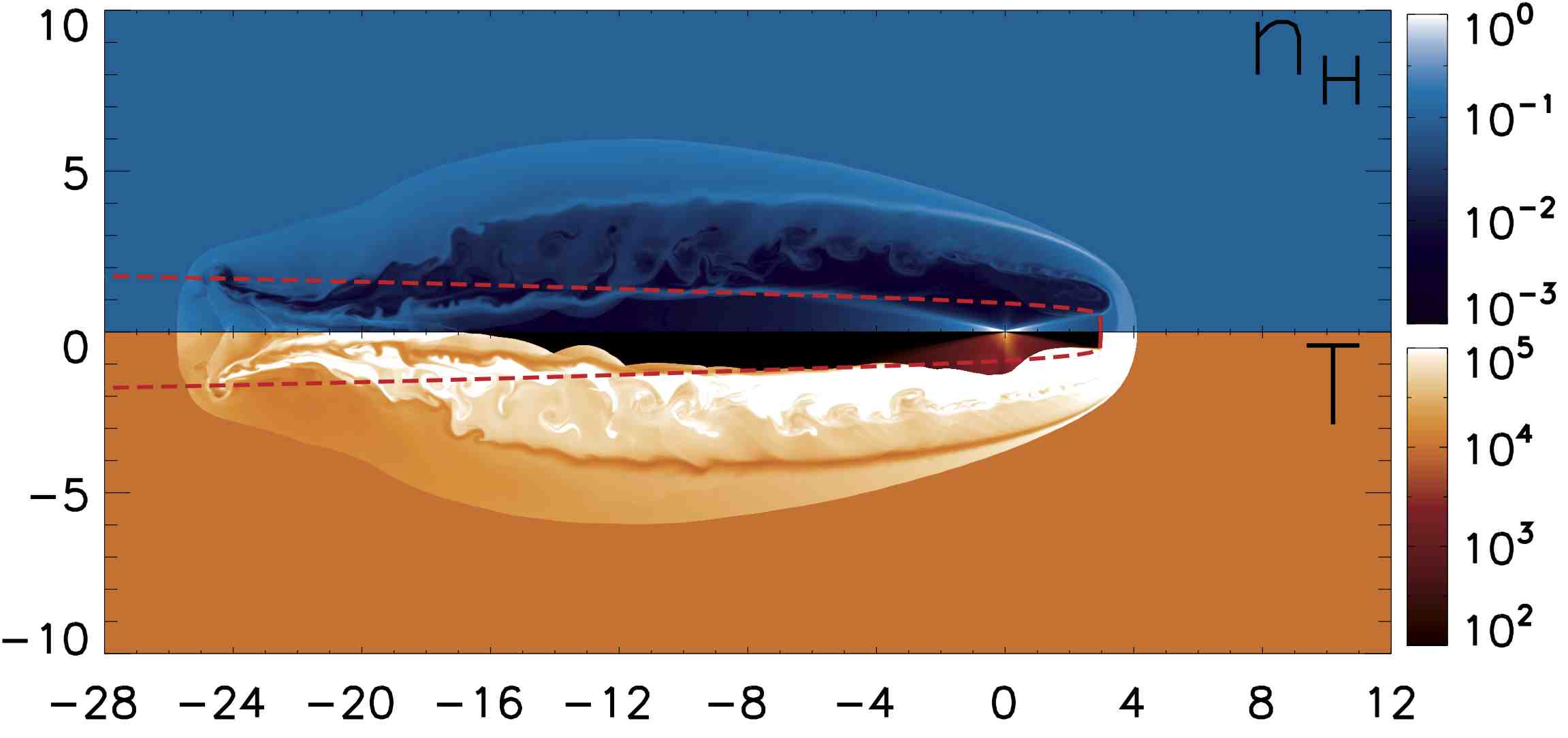}
\includegraphics[width=1.0\columnwidth]{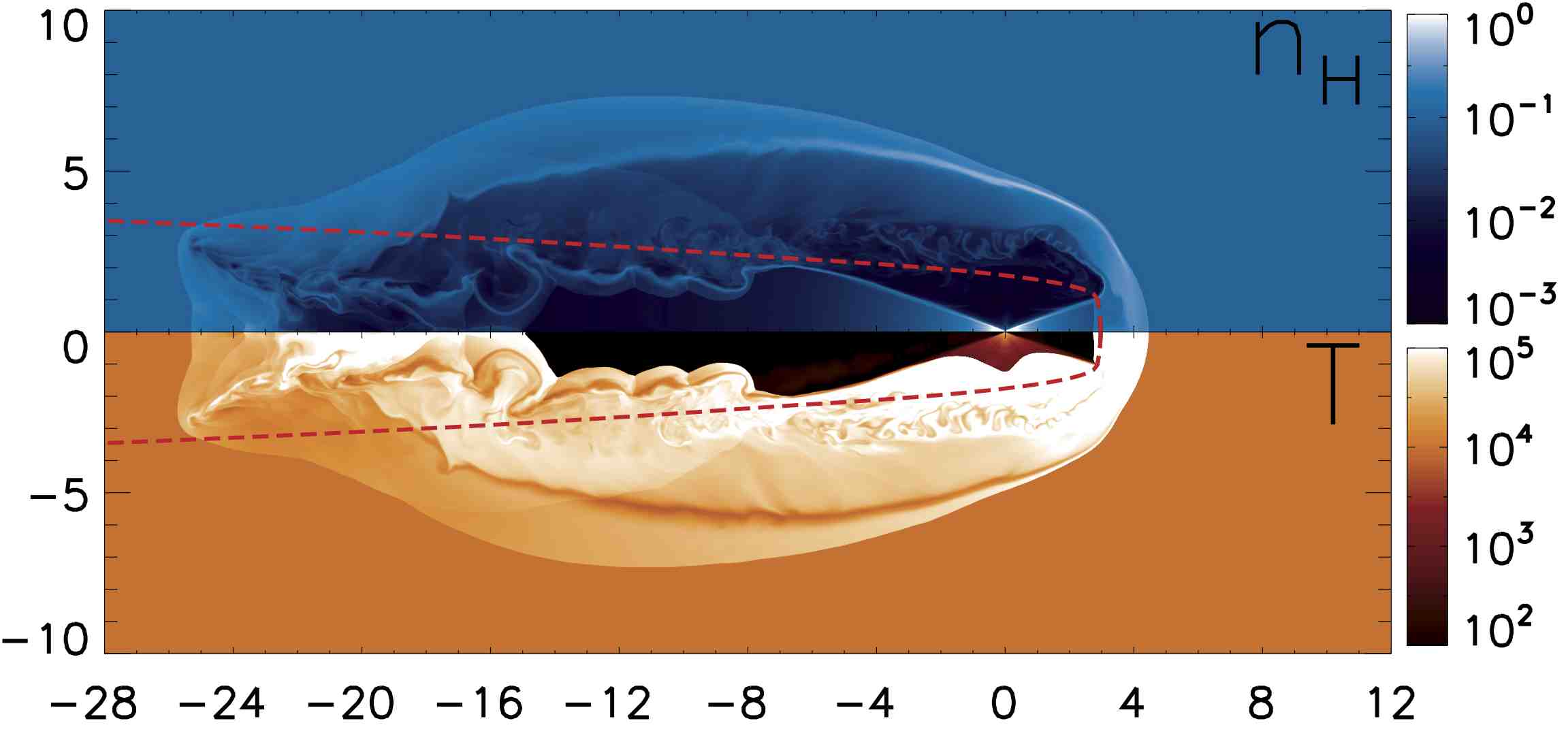}
\includegraphics[width=1.0\columnwidth]{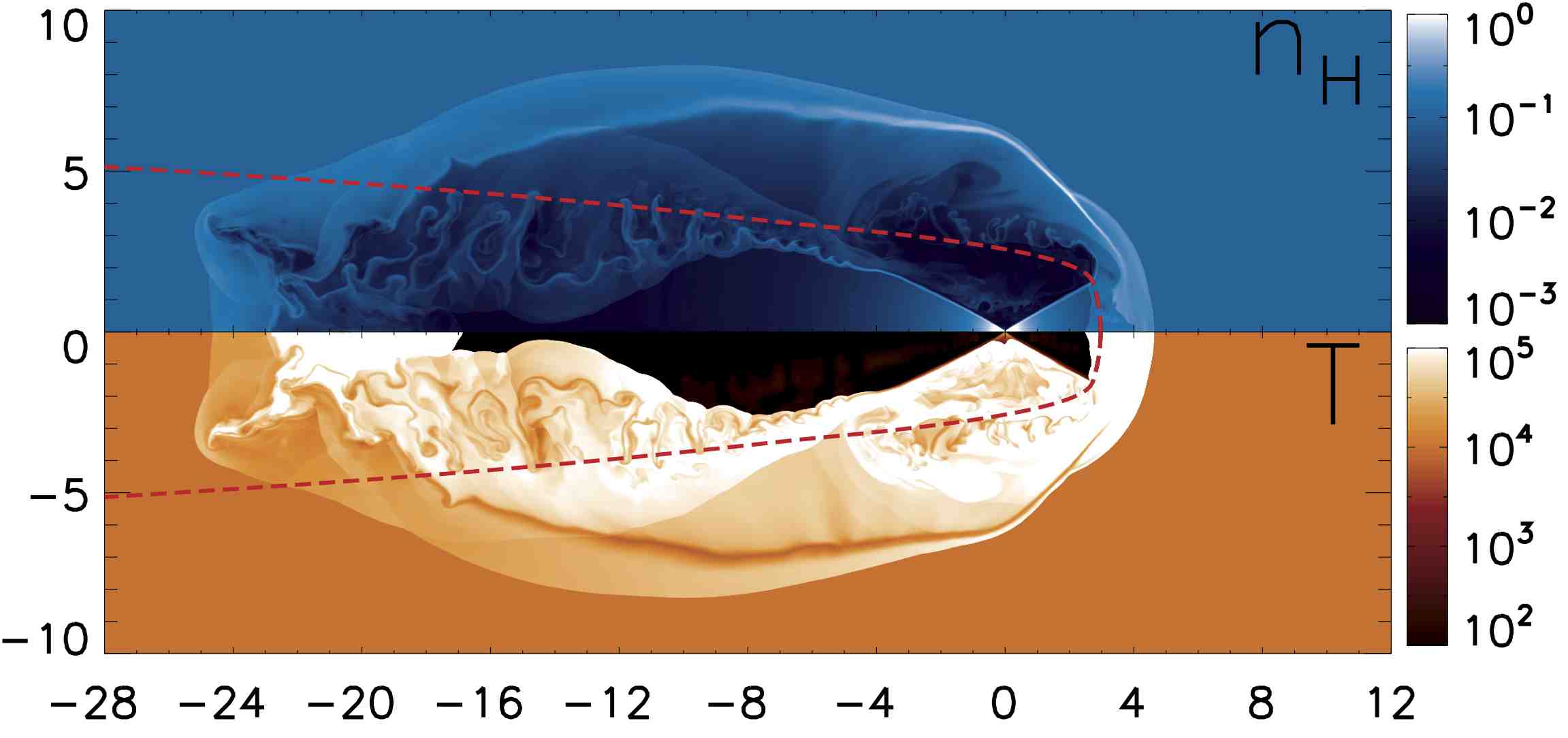}
\caption{Density stratifications (color scale given in cm$^{-3}$, blue panels); and temperature (color scale given in K, orange panels). Top, middle and bottom figures correspond to models 1, 2 and 3 (respectively). The integration time is t=5$\times10^{3}$~yrs (top). The axes are labeled in units of $10^{17}$~cm.}
\label{fig4}
\end{figure}

In Figure~\ref{fig4}, we show single time frames
($t=5\times10^{3}$~yr) of the density and temperature
stratifications (top and bottom parts, respectively, of
each of the three panels) obtained from
models 1 through 3  (see Table~1). In each plot we include the 
corresponding ``thin shell'' analytic solution
(see Section \ref{sec:anal}).

Even though models 1 through 3 differ in $\dot{M}_j$ and $\Theta$
(see Table 1), they result in basically the same flow
morphology. Model 3 (which has the largest mass loss rate and opening
angle) has the broadest flow. Independently of the $\dot{M}_j$ and
$\Theta$ values, all three models coincide with the theoretical stagnation
point ($R_0=2.97\times10^{17}$~cm), and they also have tails
with the same basic morphology. It is clear that the numerical
simulations produce broad interaction regions,
which are a direct result of the fact that the post-shock cooling
distances (${d_c}^{r}$, and ${d_c}^{f}$, see Table 1)
are not small compared to the width of the flows.
We find that the thin layer analytic solution lies along a line dividing
the computed flows into two regions: an internal low-density, hot region
(corresponding to the jet cocoon) and a high-density external,
cold region (corresponding to the shocked, streaming ambient medium).

The fact that the cooling distances are large compared to the cross
section of the jet indicates that the flows are non-radiative.
In this case it is remarkable that the analytical solution does
successfully divide the hot/low-density from the cold/high-density
regions. This is not necessarily expected, as the analytical model
assumes that the material is well mixed in a thin layer, which would
arise more naturally in radiative flows (where cooling occurs in a
small region).

\begin{figure}[!h]
\includegraphics[width=1.0\columnwidth]{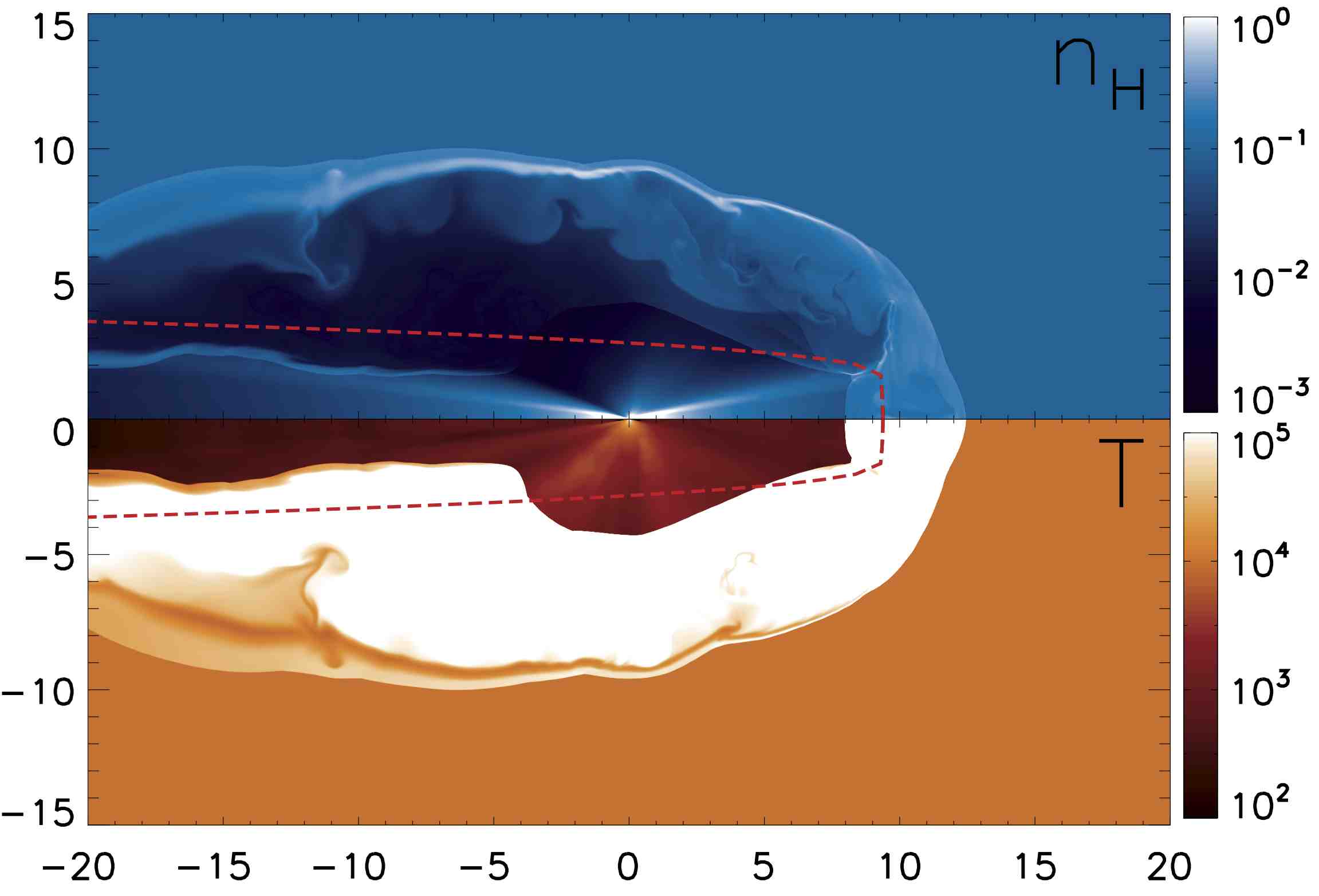}
\includegraphics[width=1.0\columnwidth]{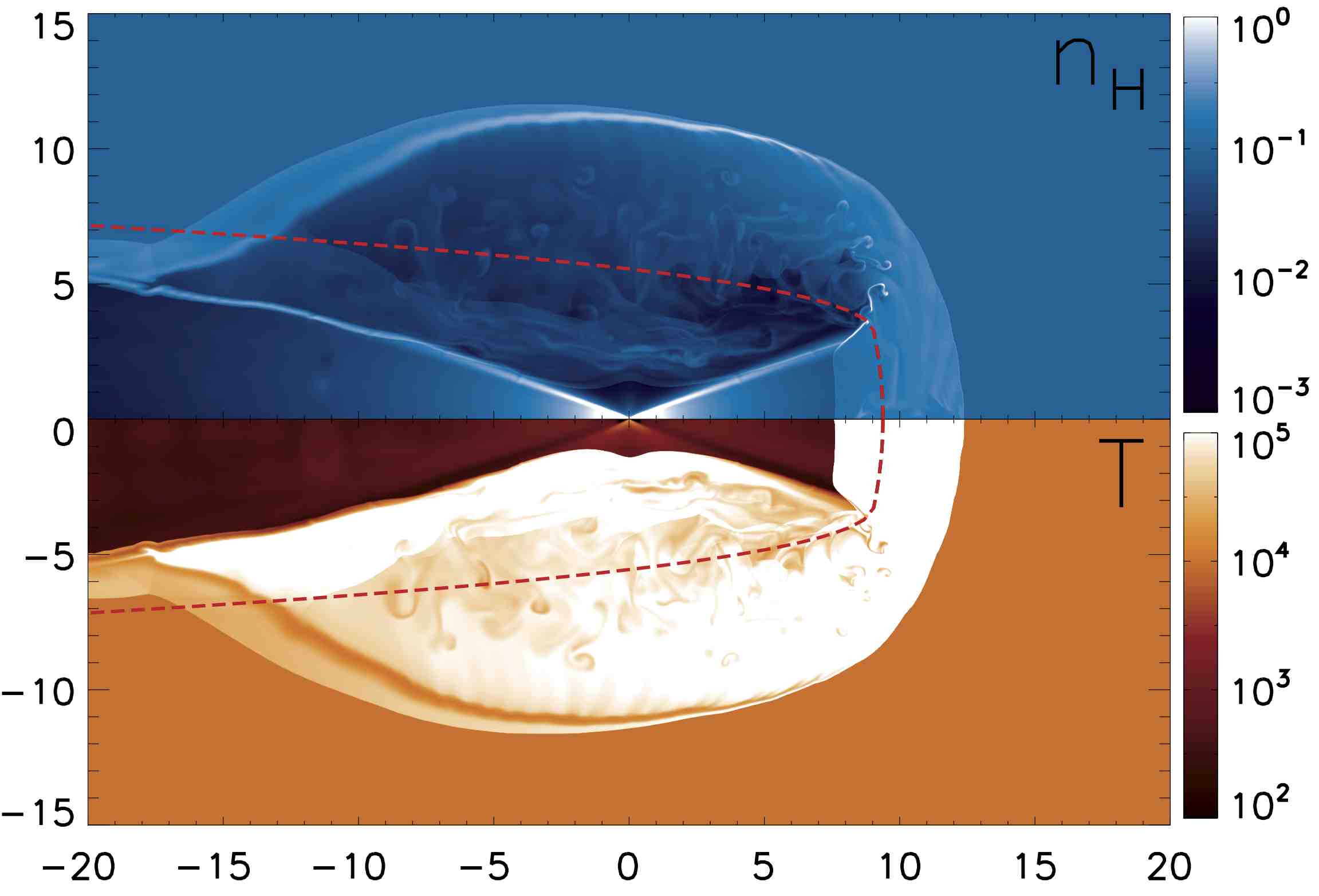}
\includegraphics[width=1.0\columnwidth]{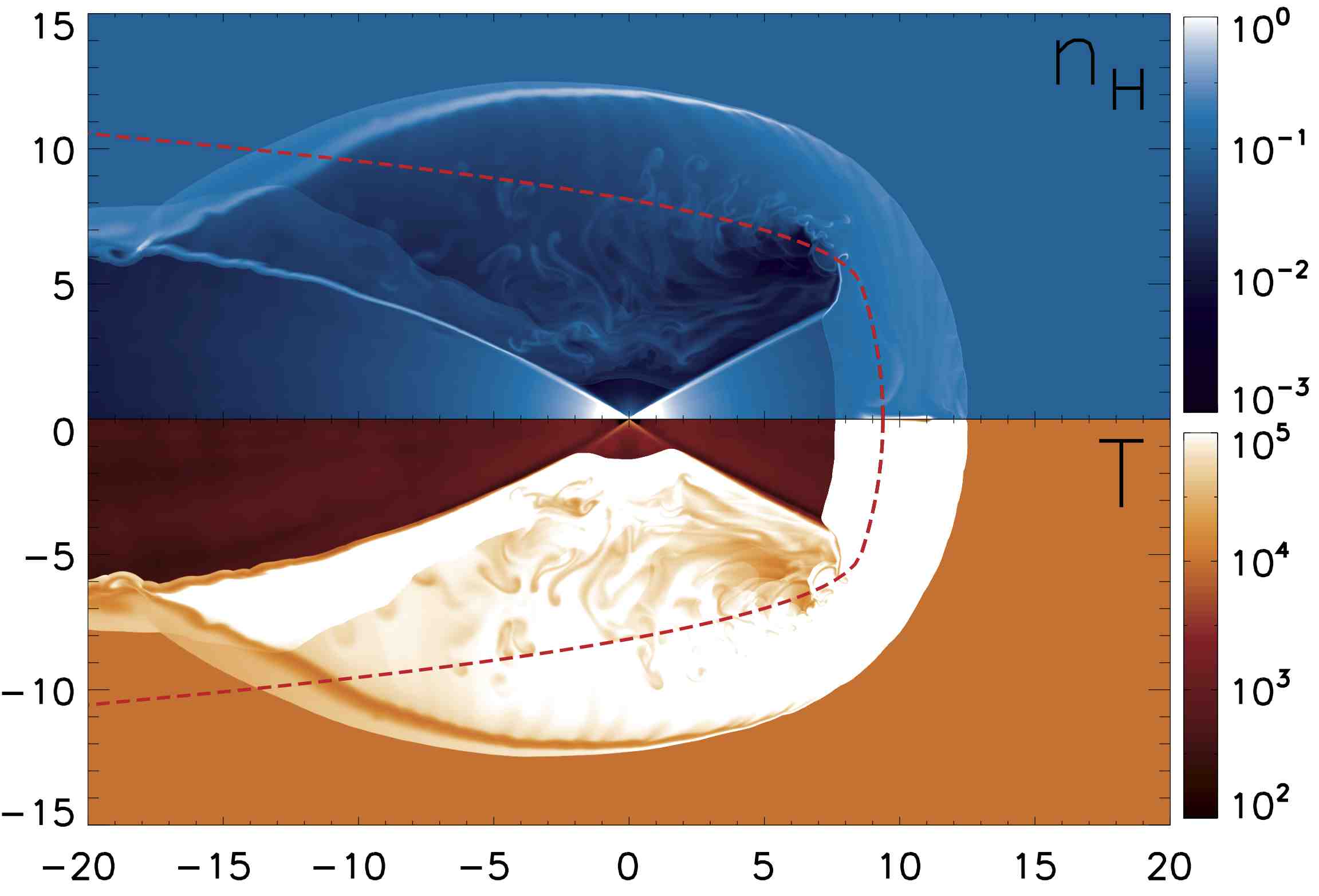}
\caption{Same as Figure~\ref{fig4} but for models 4, 5, and 6 (top, middle, and bottom respectively).}
\label{fig5}
\end{figure}

The density and temperature stratifications (at $t=5\times10^{3}$~yr)
for models 4-6 are shown in Figure~\ref{fig5}.
These models, with mass loss rates one order of
magnitude larger than those from the first three models ($\sim
10^{-9}$ M$_{\odot}$s$^{-1}$, for details see Table 1) have smaller
cooling distances than models 1-3, however they are still in the
non-radiative regime. 
The three models roughly match the analytically derived stagnation point
($R_0=9.39\times10^{17}$~cm), and produce broader flows
for larger values of the opening angle $\Theta$.

\begin{figure}[!h]
\includegraphics[width=1.0\columnwidth]{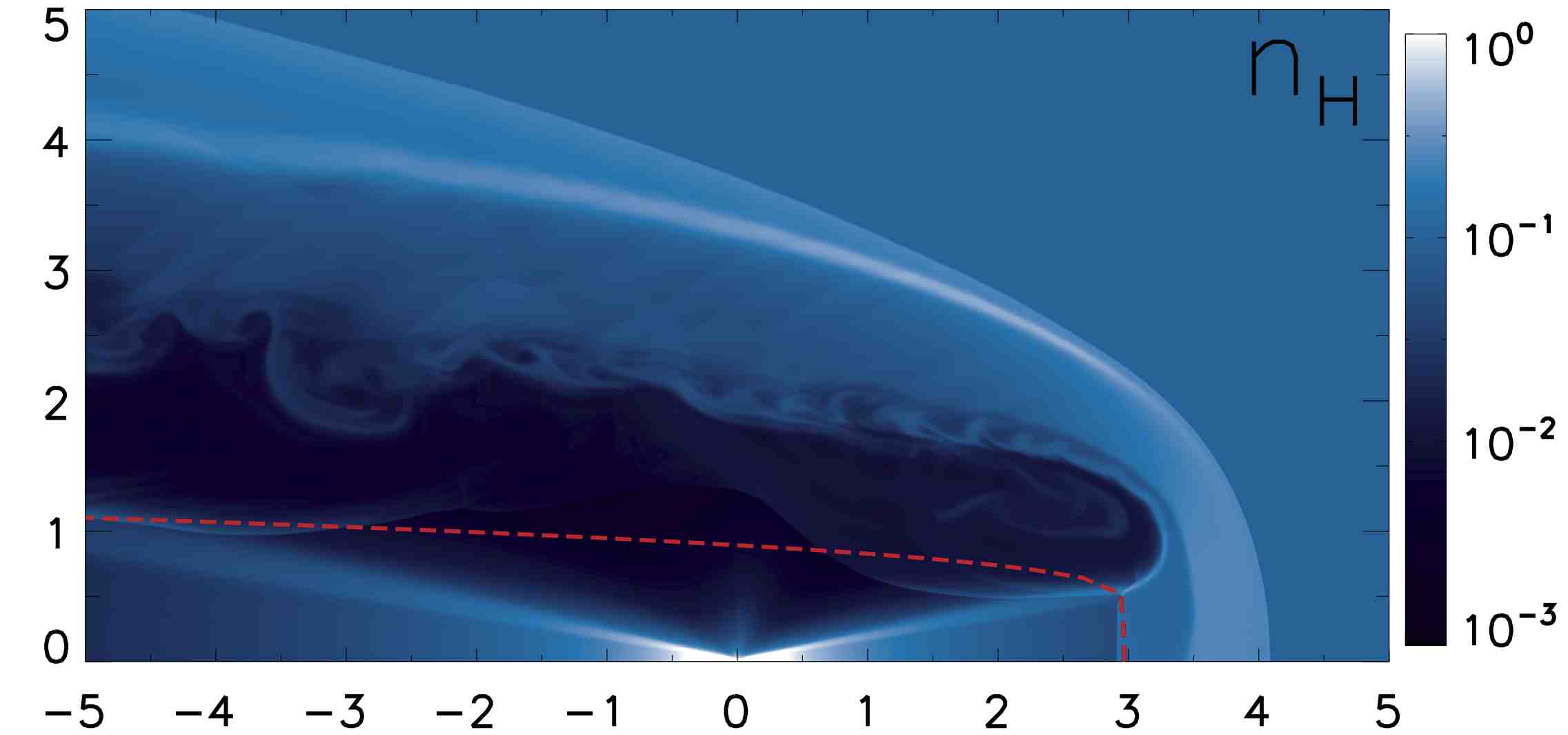}
\includegraphics[width=1.0\columnwidth]{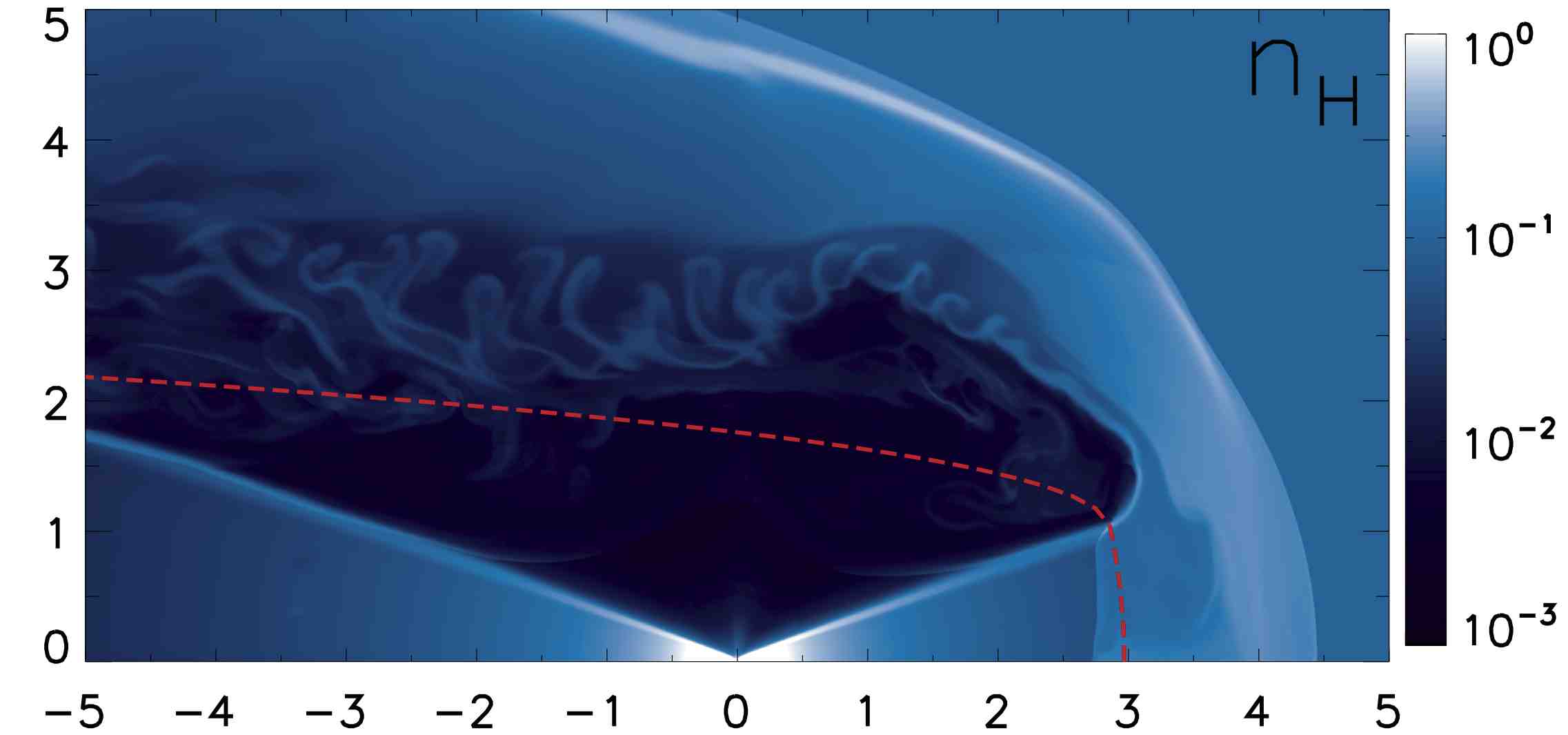}
\includegraphics[width=1.0\columnwidth]{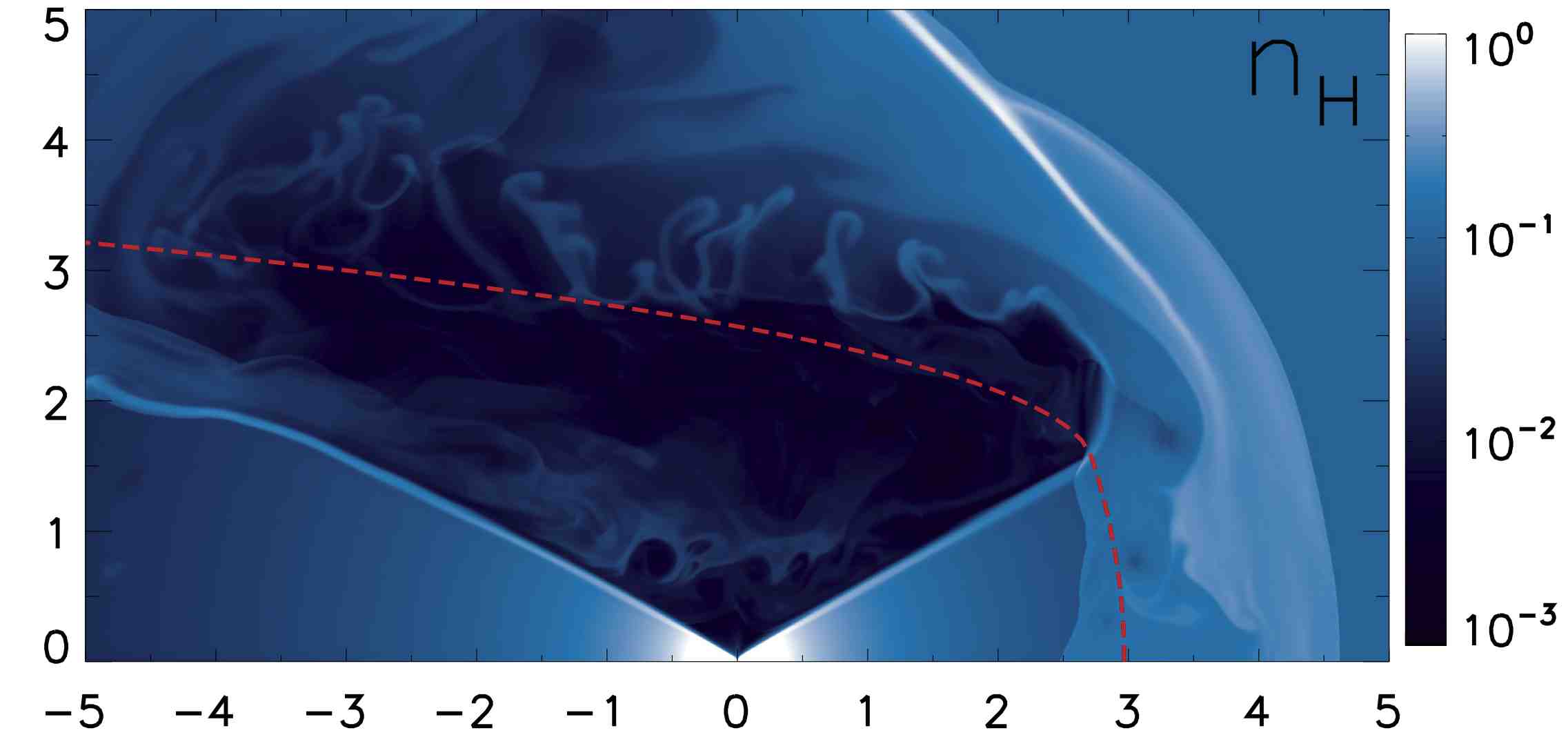}
\caption{Density stratifications (color scale given in cm$^{-3}$). Top, middle and bottom figures correspond to model 1 (and 2); 3; and 4 (respectively). The integration time is t=5$\times10^{3}$~yrs (top). The axes are labeled in units of $10^{17}$~cm. The broken line delimits the analytic solution proposed in Section~\ref{sec:anal}.}
\label{fig6}
\end{figure}

\begin{figure}[!h]
\includegraphics[width=1.0\columnwidth]{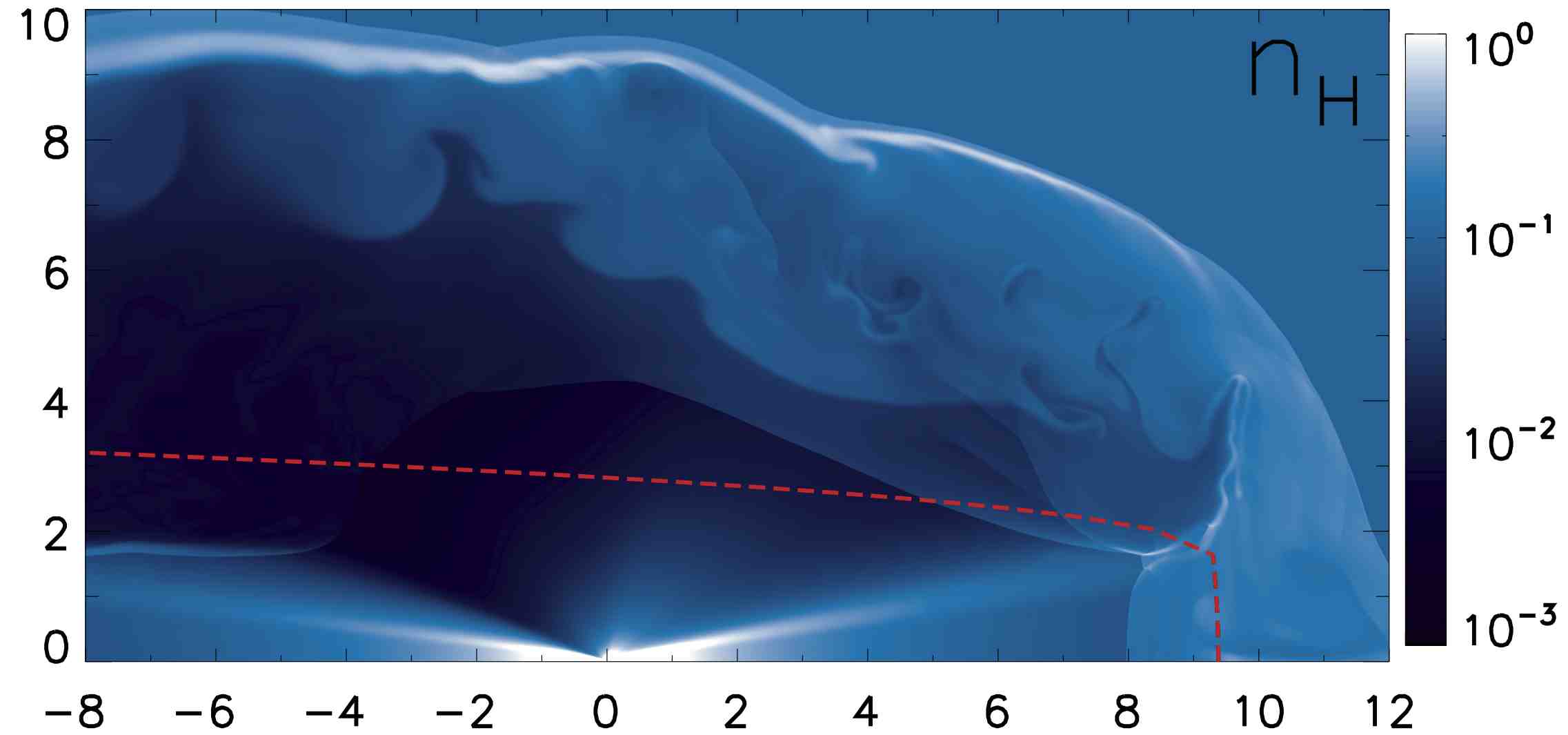}
\includegraphics[width=1.0\columnwidth]{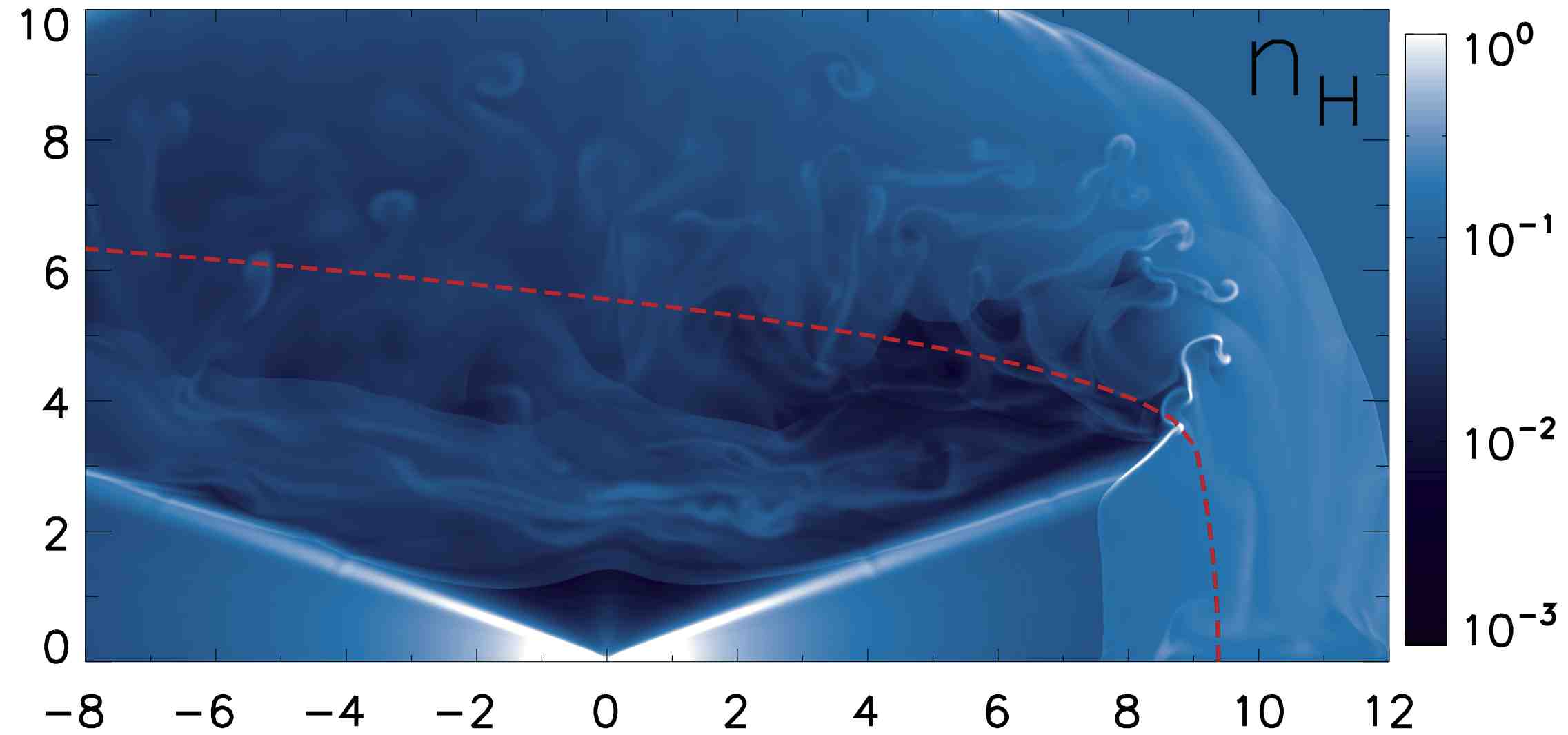}
\includegraphics[width=1.0\columnwidth]{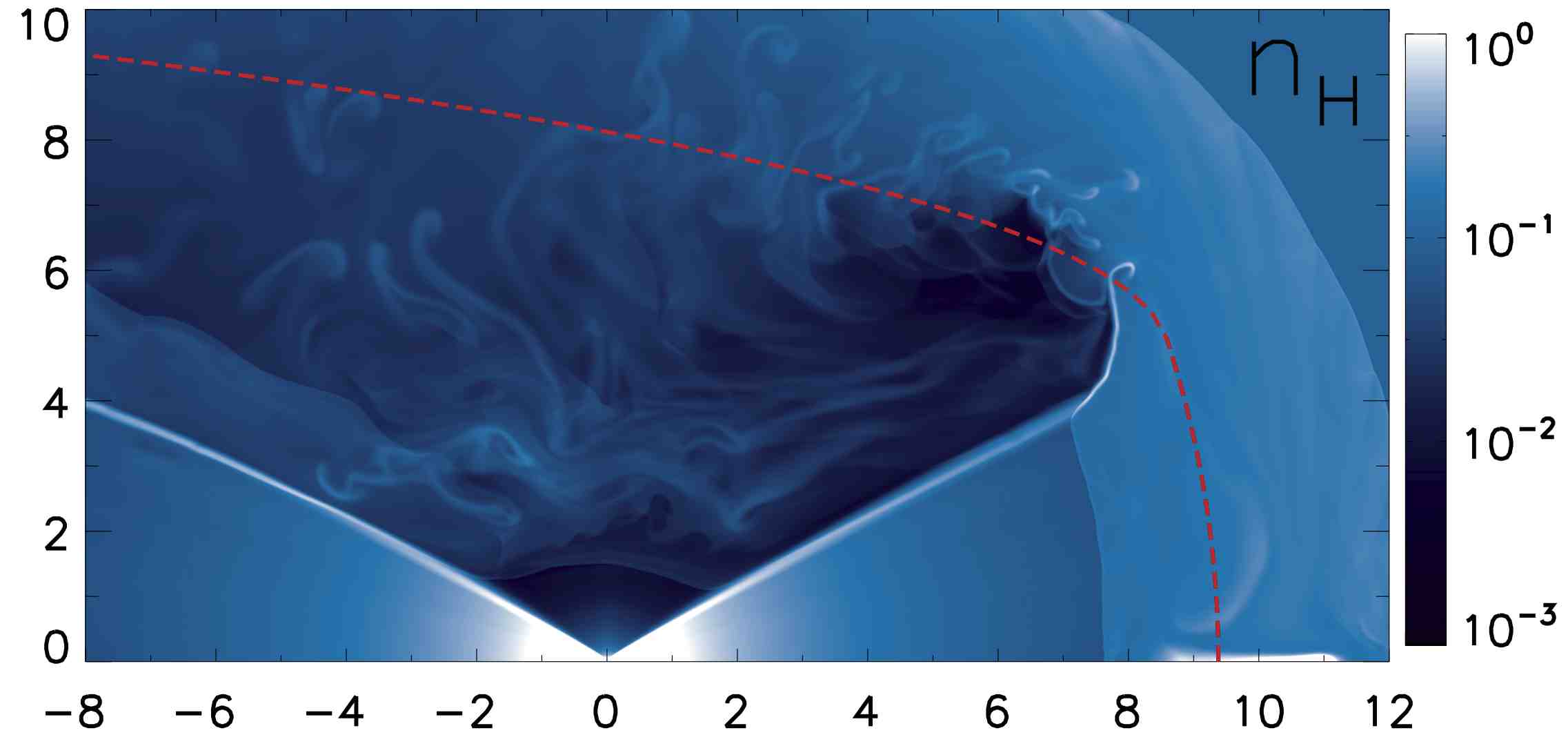}
\caption{Same as Figure~\ref{fig6} but for models 5; 6; and 7 (top, middle and bottom respectively).}
\label{fig7}
\end{figure}

In order to check how well the numerical simulations reproduced the
theoretical stagnation point in the thin shell analytical solution
from Section \ref{sec:anal}, we show a zoom of the density
structures of the jet heads for all models in Figures~\ref{fig6}-\ref{fig7}.
We find that a steady state is reached (with small discrepancies, more
evident in models 4-7, Fig. \ref{fig7}), with a shock structure approximately
centered on the stagnation radius $R_0$ (see Table
1). One can also see that the place at which the solution curves
sharply coincides nicely with the end of the dense wall of the jet.

\begin{figure}[!h]
\includegraphics[width=1.0\columnwidth]{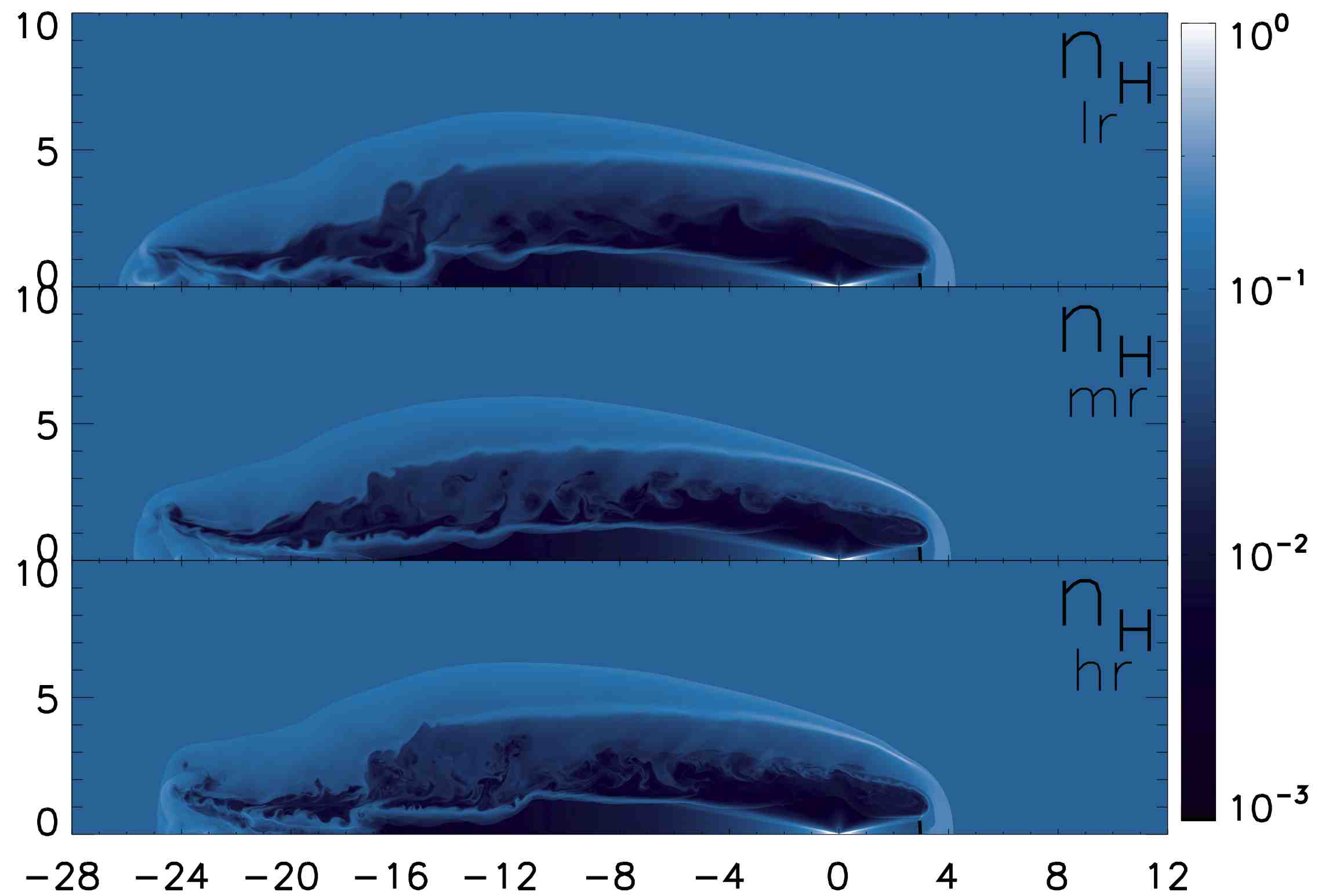}
\caption{Density stratifications (color scale given in
  cm$^{-3}$) for model 1 at t=3$\times10^{3}$~yrs, for an ``hr''
  resolution (top panel); ``mr'' resolution (middle panel); and ``lr''
  resolution (bottom panel). The axes are labeled in units of $10^{17}$~cm.}
\label{fig8}
\end{figure}

In order to illustrate the effect of the numerical resolution, we
computed two extra simulations with the setup of model
1. The resulting density stratifications for an integration time
$t=5\times10^{3}$~yr, are shown in Figure~\ref{fig8}.

Even though more complex structures are obtained for increasing resolutions, the basic morphology of the flow (stagnation point, cocoon size, density profile, etc.) is the same in all cases. Taking the number of grid points across the jet as an estimate of the Reynolds number of the simulation (Re), in these models we only reach Re~$\sim R_j / \Delta x \approx 10^{2}$ (in our hr run). To avoid Reynolds number dependency, the simulations should have resolutions larger by a factor of at least two orders of magnitude greater, which is presently unattainable. However, from the results shown in Figure 8, it is clear that the mr resolution appears to produce the correct basic morphology for the flow.
We must note that this result is similar to the one found by \citet{wzo07}. In this, the authors simulated the interaction of an AGB star moving through the ISM (with also Re~$\sim 10^{2}$). The key result was that independently of the Reynolds number, the ISM ram pressure stripped material from the head of the AGBÕs bow shock. Such stripping as well as the presence of Rayleigh-Taylor instabilities produced turbulent clump structures that eventually moved down stream to the tail of the bow shock. This mechanism even though in smaller scale, is also present in our simulations (see Figures~\ref{fig3a} - \ref{fig3c}).

\begin{figure}[!h]
\includegraphics[width=1.0\columnwidth]{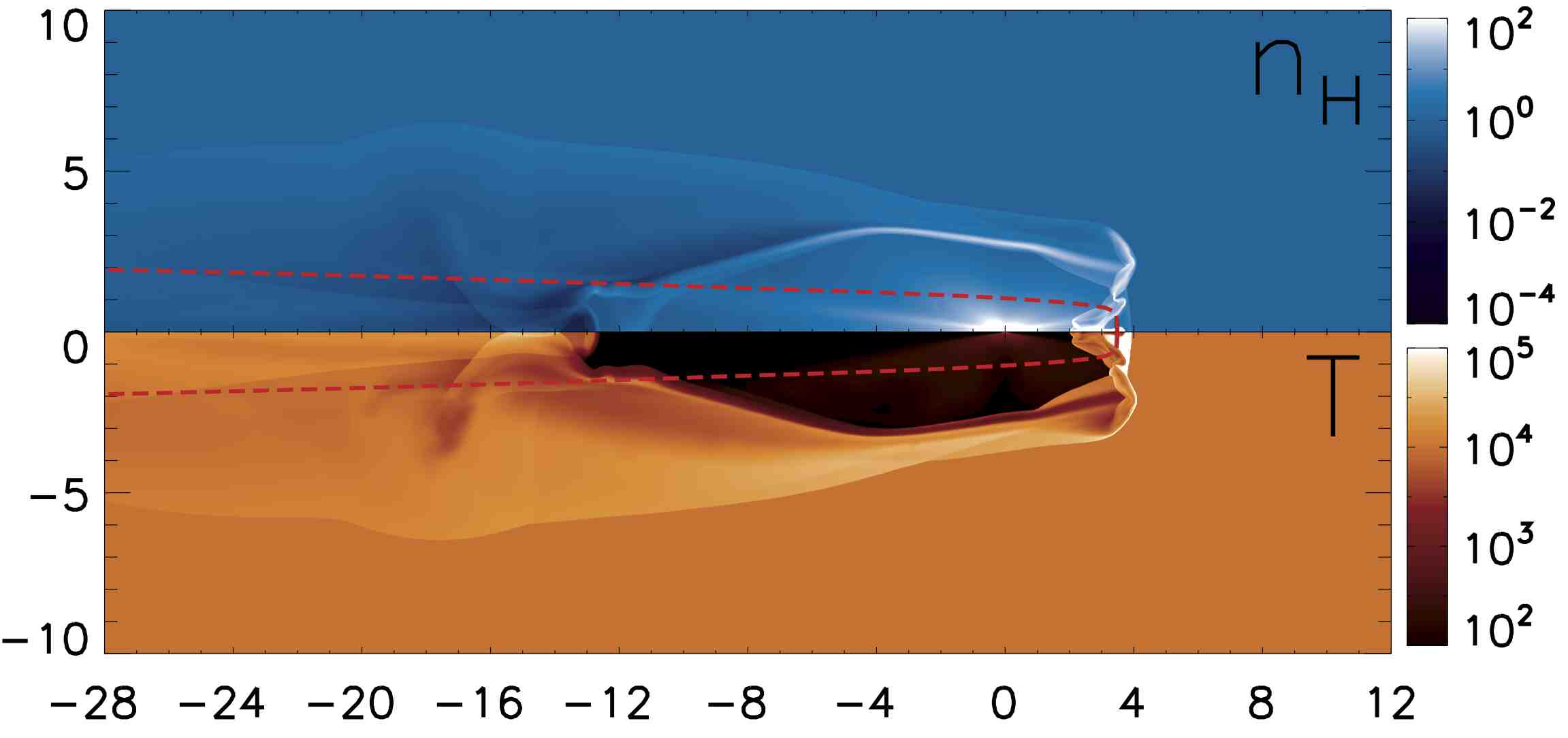}
\includegraphics[width=1.0\columnwidth]{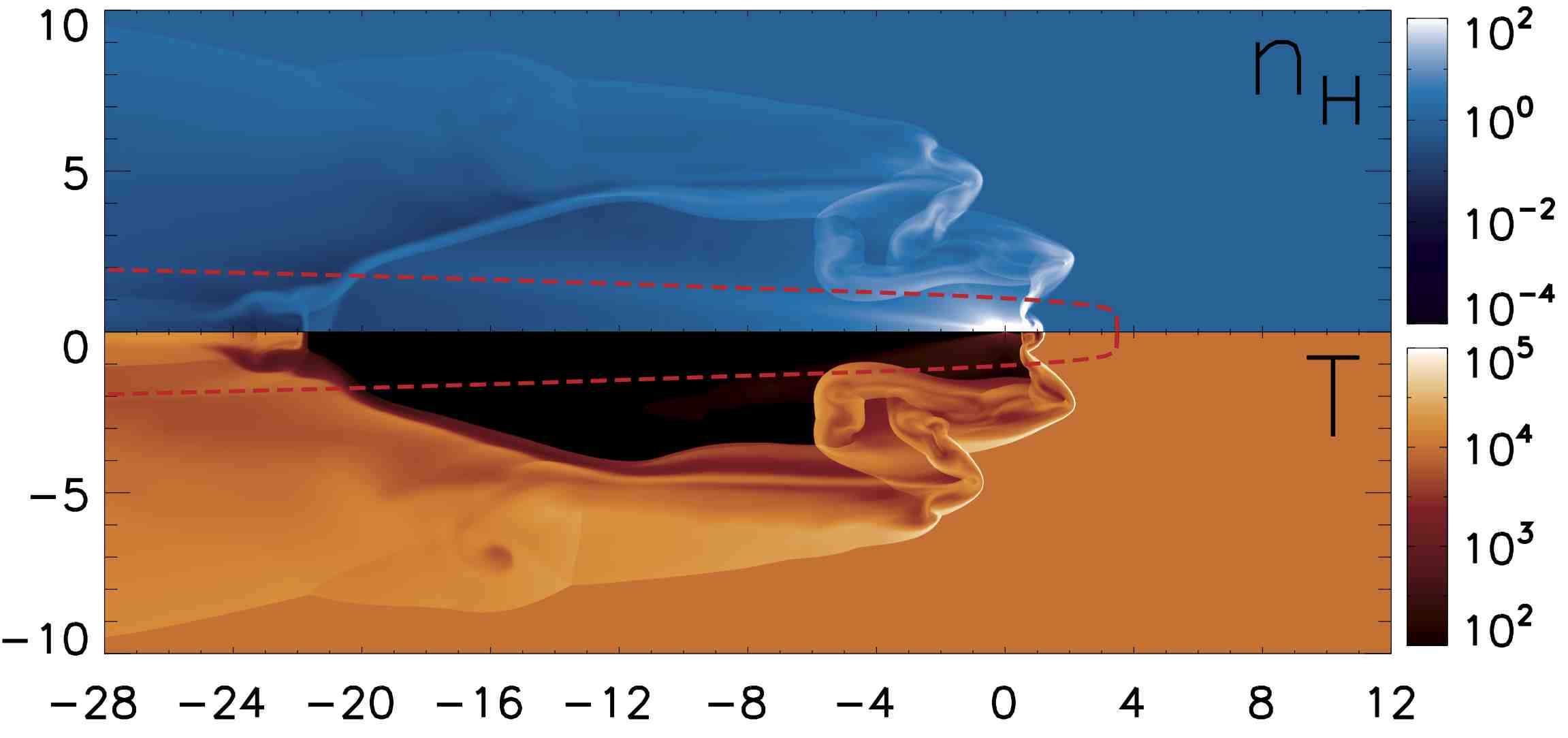}
\includegraphics[width=1.0\columnwidth]{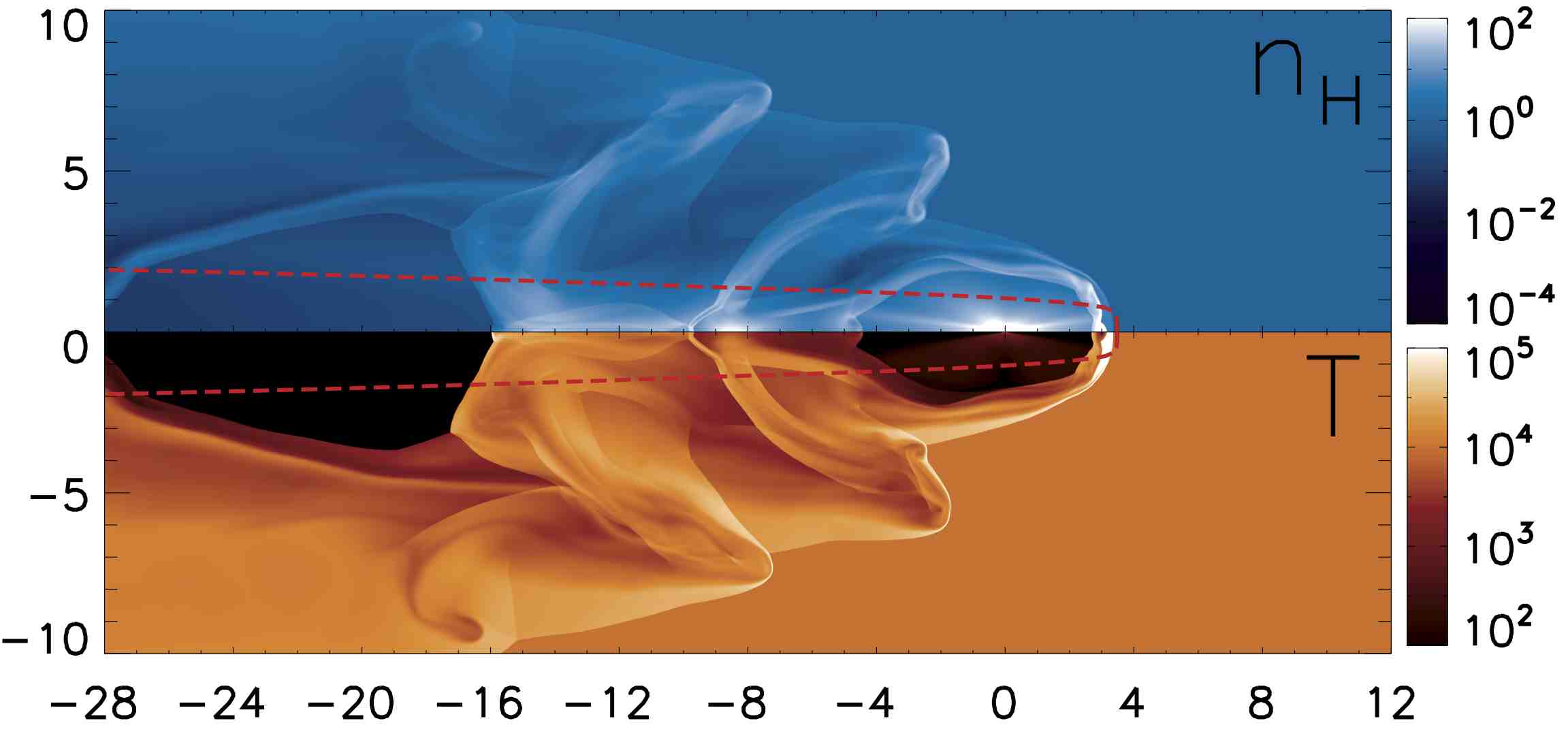}
\caption{Density stratifications (color scale given in cm$^{-3}$, blue
  panels); and temperature (color scale given in K, orange panels) of
  model 7 at t=10$\times10^{3}$~yrs (top); t=18$\times10^{3}$~yrs (middle); and t=24$\times10^{3}$~yrs (bottom). The axes are labeled in units of $10^{16}$~cm.}
\label{f9}
\end{figure}

As we have mentioned earlier, the analytic model assumes that the material of
the environment and the jet mix well in a thin shell. Such a situation
is more easily met in radiative flows, characterized by a thin cooling
layer. However, our simulations with parameters appropriate for
the jet of Mira B are basically non-radiative.
In order to check what happens with a more radiative flow
we ran model 7.
In this model, the reverse shock is indeed radiative
(${d_c}^{r}<R_0\tan\Theta$) while the cooling distance of the forward shock is of the order of the cross section of the jet (${d_c}^{f}\sim R_0\tan\Theta$) (see Table 1). This model has a 10 times smaller
computational domain, and therefore a 10 times larger resolution
than the mr models (see Section 3). In Figure~\ref{f9}, we overlap the thin layer analytic solution on top of the density and temperature stratifications obtained from model 7 at various times ($v_a=100$~km~s$^{-1}$, and
$v_j=40$~km~s$^{-1}$). For this model the stagnation point is $R_0
=3.48 \times 10^{16}$~cm, and the cooling 
distances are ${d_c}^{r} = 0.28\,R_0\tan\Theta$, and ${d_c}^{f} =
1.31\,R_0\tan\Theta$.
As can be seen from the time-sequence, a steady state is not reached
in this model. The model develops a turbulent structure which is not
stable, and the distance from the jet source to the bow-shock
oscillates around $R_0$.

\section{Discusion}
\label{sec:disc}

Motivated by the recently observed bipolar jet from Mira \citep{m09}, we
study the interaction of a steady bi-conical outflow with a streaming
environment. We use this scenario to perform 2D axisymmetrical
simulations and an analytical model of such an interaction.

We have studied the particular case in which the bi-conical jets are
aligned with the streaming environment (as shown in
Figure~\ref{fig1}). This of course is an idealization of the situation
which could possibly be found in Mira, which is unlikely to have a
perfect alignment between the outflow axis and the direction of Mira's
motion. It allows, however, simple axisymmetrical analytical and numerical
solutions which are useful as a first exploration of the more complex,
less symmetric problem of the jet in Mira B. 

The proposed ``thin shell'' analytical solution, based on the momentum
equilibrium between the outflow material from the conical jet and the
ambient medium, traces the shape of the shock (see
Equations~7-9). The model assumes that the interaction occurs in a
very thin layer in which the material from the jet and the ISM mixes
well. This situation is likely to occur in radiative flows, there the
size of the cooling distances are small compared with the flow
dimensions. This, however is not the case of the resulting interaction
of the outflow launched from Mira B and its surrounding environment.

The models reach a stationary state with a fixed stand-off distance between
the jet head and the outflow source. This configuration might
be relevant for the collimated outflow from Mira, in which the
forward directed jet head has a very low proper motion with
respect to Mira (see \citet{m09}).

We find that the non-radiative flow configurations (that result
from the parameters deduced from the bipolar outflow from Mira)
have broad interaction regions which approximately coincide with
the predictions from the analytic model (in which a thin shell interaction
is assumed). Also, the analytical solutions lie close
to the contact discontinuity dividing the material in the jet cocoon
from the shocked environment region.

We have tested the convergence of the numerical simulations
for increasing resolutions. We find that at different resolutions
we obtain similar large-scale flow structures, but with different
structures at smaller scales (in which the Kelvin-Helmholtz instabilities
associated with shear layers are active). Better convergence is expected
only for resolutions higher by at least two orders of magnitude, in which
the simulations would start to approach the ``high Reynolds number regime''
appropriate for the real, astrophysical flow.

We end by noting again that the present model of a bipolar jet/streaming ISM interaction is not completely consistent with the outflow from the Mira system \citep{m09}. This is because in the binary Mira AB system, the mass loss rate from Mira A's uncollimated wind (with $\dot{M} \sim $10$^{-6} M_{\odot}$s$^{-1}$, \citet{ir07}) clearly dominates the interaction with the ambient medium. The mass loss from the bi-conical jet ($\dot{M_j} \sim $10$^{-9} M_{\odot}$s$^{-1}$, inferred from \citet{m07}), will simply modify the morphology created by the interaction from the isotropic wind and the ambient medium, except in the region in which the leading jet from Mira B emerges from the Mira A wind bubble. This is the region to which our model would in principle apply.

Future models of the outflow from Mira should include both the
wind from the primary star and the collimated outflow from the
secondary. This more complex flow will clearly be the topic of
future papers on Mira's outflow/ISM interaction.

\section*{Acknowledgments}
DLC work has been possible thanks to a DGAPA postdoctoral grant from the UNAM.  
We also acknowledge support from CONACyT grants 61547, 101256, and
101975.  
We thank Enrique Palacios and Mart\'\i n Cruz for supporting the
servers in which the calculations of this paper were carried out.

\end{document}